\documentstyle[11pt,paspconf]{article}
\newcommand{\mincir}{\raise -2.truept\hbox{\rlap{\hbox{$\sim$}}\raise5.truept
\hbox{$<$}\ }}
\newcommand{\magcir}{\raise -2.truept\hbox{\rlap{\hbox{$\sim$}}\raise5.truept
\hbox{$>$}\ }}
\newcommand{\minmag}{\raise-2.truept\hbox{\rlap{\hbox{$<$}}\raise 6.truept\hbox
{$>$}\ }}
\newcommand{\be}{\begin{equation}}
\newcommand{\ee}{\end{equation}}
\newcommand{\ba}{\begin{eqnarray}}
\newcommand{\ea}{\end{eqnarray}}
\newcommand{\brr}{\begin{array}}
\newcommand{\err}{\end{array}}
\newcommand{\bc}{\begin{center}}
\newcommand{\ec}{\end{center}}

\begin{document}
\title{Lagrangian Dynamics of Collisionless Matter}

\author{Sabino Matarrese$^1$ \& David Terranova$^1$}
\affil{$^1$ Dipartimento di Fisica ``G. Galilei", Universit\`a di Padova, 
       Italy}

\begin{abstract}
The non--linear dynamics of self--gravitating irrotational
dust is analyzed in a general relativistic framework, using synchronous and
comoving coordinates. Writing the equations in terms of the metric 
tensor of the spatial sections orthogonal to the fluid flow 
allows an unambiguous expansion in inverse powers of the speed of light.
The Newtonian and post--Newtonian approximations are derived in 
Lagrangian form. A general formula for the gravitational 
waves generated by the non--linear evolution of cosmological perturbations
is given. 
It is argued that a stochastic gravitational--wave background is produced by 
non--linear cosmic structures, with present--day closure density $\Omega_{gw} 
\sim 10^{-5}$ -- $10^{-6}$ on Mpc scale.
\end{abstract}

\section{Introduction}

The gravitational instability of collisionless matter in a cosmological 
framework is usually studied within the Newtonian approximation, 
which basically consists in neglecting terms higher than the first in 
metric perturbations around a matter--dominated Friedmann--Robertson--Walker 
(FRW) background, while keeping non--linear density and velocity perturbations. 
This approximation is usually thought to produce accurate 
results in a wide spectrum of cosmological scales, namely on 
scales much larger than the Schwarzschild radius of collapsing bodies and 
much smaller than the Hubble horizon scale, where the peculiar gravitational 
potential $\varphi_g$, divided by the square of the speed of light $c^2$ to 
obtain a dimensionless quantity, keeps much less than unity, while the peculiar 
matter flow never becomes relativistic. 
To be more specific, the Newtonian approximation 
consists in perturbing only the time--time component of the FRW 
metric tensor by an amount $2\varphi_g/c^2$, where $\varphi_g$ 
is related to the matter density fluctuation $\delta$ via the 
cosmological Poisson equation, 
$\nabla_x^2 \varphi_g ({\vec x},\tau) = 4 
\pi G a^2(\tau) \varrho_b(\tau) \delta({\vec x}, \tau)$, 
where $\varrho_b$ is the background matter density, 
$a(\tau)$ the appropriate FRW scale--factor and $\tau$ the conformal time. 
The fluid dynamics is then usually studied in Eulerian coordinates by 
accounting for mass conservation 
and using the cosmological version of the Euler equation for a 
self--gravitating pressureless fluid to close the system. 
To motivate the use of this ``hybrid approximation", which 
deals with perturbations of the matter and the geometry at a different 
perturbative order, one can either formally expand the correct 
equations of General Relativity (GR) in inverse powers of the speed of light
or simply notice that the peculiar gravitational potential is strongly 
suppressed with respect to the matter perturbation by the square of the ratio 
of the perturbation scale $\lambda$ to the Hubble radius $r_H= c H^{-1}$ 
($H$ being the Hubble constant): 
$\varphi_g/c^2 \sim \delta ~(\lambda / r_H)^2$. 

Such a simplified approach, however, already fails in producing an accurate 
description of the trajectories of relativistic particles, such as photons. 
Neglecting the relativistic perturbation of the space--space components 
of the metric, which in the so--called longitudinal gauge is just 
$-2\varphi_g/c^2$, would imply a mistake by a factor of two in 
well--known effects such as the Sachs--Wolfe, Rees--Sciama and 
gravitational lensing. The level of accuracy not only depends on the peculiar 
velocity of the matter producing the spacetime curvature, but also on the 
nature of the particles carrying the signal to the observer. 
Said this way, it may appear that the only relativistic correction required 
to the usual Eulerian Newtonian picture is that of writing the metric tensor 
in the ``weak field" form (e.g. Peebles 1993)
\be 
ds^2 = a^2(\tau) \biggl[ - \biggl(1 + {2\varphi_g \over c^2} \biggr) ~c^2 
d\tau^2 + \biggl(1 - {2\varphi_g \over c^2} \biggr) ~d l^2 \biggr] \;.
\ee

As we are going to show, this is not the whole story. 
It is well--known in fact that 
the gravitational instability of aspherical perturbations (which is the generic 
case) leads to the formation of very anisotropic structures whenever pressure 
gradients can be neglected (e.g. Shandarin et al. 1995 and references therein). 
Matter first flows in almost two--dimensional structures called pancakes, 
which then merge and fragment to eventually form one--dimensional filaments 
and point--like clumps. 
During the process of pancake formation the matter density, the shear 
and the tidal field formally become infinite along evanescent 
two--dimensional configurations corresponding to caustics; after this
event a number of highly non--linear phenomena, such as vorticity 
generation by multi--streaming, merging, tidal disruption and 
fragmentation, occur. 
Most of the patology of the caustic formation process, such as the local 
divergence of the density, shear and tide, and the formation of multi--stream 
regions, are just an artifact of extrapolating the pressureless 
fluid approximation beyond 
the point at which pressure gradients and viscosity become important. 
In spite of these limitations, however, it is generally believed that 
the general anisotropy of the collapse configurations, either pancakes or 
filaments, is a generic feature of cosmological structures originated through 
gravitational instability, which would survive even in the presence of a
collisional component. 

This simple observation shows the inadequacy of the standard Newtonian 
paradigm. According to it the lowest scale at which the approximation can 
be reasonably applied is set by the amplitude of the gravitational potential 
and is given by the Schwarzschild radius of the collapsing body, which is 
negligibly small for any relevant cosmological mass scale. 
What is completely missing in this criterion is the role of the shear which 
causes the presence of non--scalar contributions to the metric perturbations. 
A non--vanishing shear component is in fact an unavoidable feature of 
realistic cosmological perturbations and affects the dynamics 
in (at least) three ways, all related to non--local effects, i.e. to the 
interaction of a given fluid element with the environment. 
First, at the lowest perturbative order the shear is related to the 
tidal field generated by the surrounding material by a simple proportionality 
law. Second, it is related to a {\em dynamical} tidal induction: the 
modification 
of the environment forces the fluid element to modify its shape and density. 
In Newtonian gravity, this is an {\em action--at--a--distance} effect, which 
starts to manifest itself in second--order perturbation theory as an 
inverse--Laplacian contribution to the velocity potential (e.g. Catelan et al. 
1995). 
Third, and most important here, a non--vanishing shear field leads to the 
generation of a traceless and divergenceless metric perturbation which can be 
understood as gravitational radiation emitted by non--linear perturbations. 
This contribution to the metric perturbations is statistically 
small on cosmologically interesting scales, but it becomes relevant whenever 
anisotropic (with the only exception of exactly one--dimensional) collapse 
takes place. In the Lagrangian picture such an effect
already arises at the post--Newtonian (PN) level. 
Note that the two latter effects are only detected if one 
allows for non--scalar perturbations in physical quantities. Contrary to a 
widespread belief, in fact, the choice of scalar perturbations in the initial 
conditions is not enough to prevent tensor modes to arise beyond the linear 
regime in a GR treatment. Truly tensor perturbations are dynamically generated 
by the gravitational instability of initially scalar perturbations, 
independently of the initial presence of gravitational waves. 
This point is very clearly displayed in the GR Lagrangian second--order 
perturbative approach. The pioneering work in this field is by 
Tomita (1967), who calculated the gravitational waves 
$\pi^\alpha_{~\beta}$ emitted by 
non--linearly evolving scalar perturbations in an Einstein--de Sitter 
background, in the synchronous gauge. Matarrese, Pantano \& Saez 
(1994a,b) obtained an equivalent result but with a different formalism
in comoving and synchronous coordinates. 

Recently a number of different approaches to 
relativistic effects in the non--linear dynamics of cosmological 
perturbations have been proposed. Matarrese, Pantano \& Saez (1993) proposed 
an algorithm based on neglecting the magnetic part of the Weyl tensor 
in the dynamics, obtaining strictly local fluid--flow evolution equations, 
i.e. the so--called ``silent universe". This formalism, however, cannot be 
applied to cosmological structure formation {\em inside} the horizon, 
where the non--local tidal  
induction cannot be neglected, i.e. the magnetic Weyl tensor 
$H^\alpha_{~\beta}$ is non--zero, with the exception of highly specific initial 
configurations (Matarrese et al. 1994a; Bertschinger \& Jain 1994;
Bruni, Matarrese \& Pantano 1995a; the dynamical role of $H^\alpha_{~\beta}$ 
was also discussed by Bertschinger \& Hamilton 1994 and Kofman \& Pogosyan 
1995). 
Rather, it is probably related to the non--linear dynamics of an irrotational 
fluid {\em outside} the (local) horizon (Matarrese et al. 1994a,b). 
One possible application (Bruni, Matarrese \& Pantano 1995b), is in fact 
connected to the {\em Cosmic No--hair Theorem}. 
Matarrese \& Terranova (1995) followed the more ``conservative" approach of 
expanding the Einstein and continuity equations in inverse powers of the 
speed of light, which then defines a Newtonian limit and, at the next 
order, post--Newtonian corrections. Their approach differs from previous ones, 
because of the gauge choice: we used synchronous and comoving coordinates, 
because of which this approach can be called a Lagrangian one. 
Various approaches have been proposed in the literature, which are somehow 
related. A PN approximation has been followed by Futamase 
(1991) to describe the dynamics of a clumpy universe. 
Tomita (1991) used non--comoving coordinates in
a PN approach to cosmological perturbations. 
Shibata \& Asada (1995) recently developed a PN approach to cosmological 
perturbations, also using non--comoving coordinates. 
Kasai (1995) analyzed the non--linear 
dynamics of dust in the synchronous and comoving gauge. 

\section{Method}

We consider a pressureless fluid with vanishing vorticity. Using
synchronous and comoving coordinates, the line--element reads 
\be
ds^2 = a^2(\tau)\big[ - c^2 d\tau^2 + \gamma_{\alpha\beta}({\vec q}, \tau) 
dq^\alpha d q^\beta \big] \;, 
\ee
where we have factored out the scale--factor of the isotropic FRW solutions. 

By subtracting the isotropic Hubble--flow, we introduce a {\em peculiar 
velocity--gradient tensor} 
$\vartheta^\alpha_{~\beta} = {1 \over 2} \gamma^{\alpha\gamma} 
{\gamma_{\gamma\beta}}'$, 
where primes denote differentiation with respect to $\tau$. 

Thanks to the introduction of this tensor we can write the Einstein's 
equations in a cosmologically convenient form. 
The energy constraint reads 
\be
\vartheta^2 - \vartheta^\mu_{~\nu} \vartheta^\nu_{~\mu} + 4 {a' \over a} 
\vartheta + c^2 \bigl( {\cal R} - 6 \kappa \bigr) = 16 \pi G a^2 
\varrho_b \delta \;,
\ee
where ${\cal R}^\alpha_{~\beta}(\gamma)$ is the 
conformal Ricci curvature of the three--space with metric 
$\gamma_{\alpha\beta}$; for the background FRW solution 
$\gamma^{FRW}_{\alpha\beta} = (1 + {\kappa\over 4} q^2)^{-2} 
\delta_{\alpha\beta}$, one has ${\cal R}^\alpha_{~\beta}(\gamma^{FRW}) 
= 2 \kappa \delta^\alpha_{~\beta}$. 
We also introduced the density contrast 
$\delta \equiv (\varrho - \varrho_b) /\varrho_b$. 

The momentum constraint reads 
\be
\vartheta^\alpha_{~\beta||\alpha} = \vartheta_{,\beta} \;. 
\ee
The double vertical bars denote covariant derivatives in the 
three--space with metric $\gamma_{\alpha\beta}$. 

Finally, after replacing the density from the energy constraint and
subtracting the background contribution, the extrinsic curvature evolution 
equation becomes 
\be
{\vartheta^\alpha_{~\beta}}' + 2 {a' \over a} \vartheta^\alpha_{~\beta} + 
\vartheta \vartheta^\alpha_{~\beta} + {1 \over 4} 
\biggl( \vartheta^\mu_{~\nu} \vartheta^\nu_{~\mu} - \vartheta^2 \biggr) 
\delta^\alpha_{~\beta} + {c^2 \over 4} \biggl[ 4 {\cal R}^\alpha_{~\beta} 
- \bigl( {\cal R} + 2 \kappa \bigr) \delta^\alpha_{~\beta} \biggr]
= 0 \;. 
\ee

The Raychaudhuri equation for the evolution of the 
{\em peculiar volume--expansion scalar} $\vartheta$ reads 
\be
\vartheta' + {a' \over a} \vartheta + \vartheta^\mu_{~\nu} \vartheta^\nu_{~\mu} 
+ 4 \pi G a^2 \varrho_b \delta =0 \;. 
\ee
The main advantage of this formalism is that there is only one dimensionless 
(tensor) variable in the equations, namely the spatial metric tensor 
$\gamma_{\alpha\beta}$. The only remaining variable is the density contrast 
which can be written in the form
\be
\delta({\vec q}, \tau) = (1 + \delta_0({\vec q})) \bigl[\gamma({\vec q}, \tau)/ 
\gamma_0 ({\vec q}) \bigr]^{-1/2} - 1 \;,
\ee
where $\gamma \equiv {\rm det} ~\gamma_{\alpha\beta}$. 

\section{Results and conclusions} 

The method is then based on a $1/c^2$ expansion of equations above which 
first of all leads to a new, purely Lagrangian, derivation of 
the Newtonian approximation (Matarrese \& Terranova 1995). One of the most 
important result in this 
respect is that we obtained a simple expression for the 
Lagrangian metric; exploiting the vanishing of the spatial 
curvature in the Newtonian limit we were able to write it in terms of the 
displacement vector ${\vec S}({\vec q}, \tau) = {\vec x}({\vec q},\tau)  - 
{\vec q}$, from the Lagrangian coordinate ${\vec q}$ to the Eulerian 
one ${\vec x}$ of each fluid element (e.g. Buchert 1995 and references 
therein), namely
\be
d s^2 = a^2(\tau) \biggl[ - c^2 d \tau^2 + \delta_{AB} 
\biggl(\delta^A_{~\alpha} + {\partial S^A({\vec q}, \tau) 
 \over \partial q^\alpha} \biggr) 
\biggl(\delta^B_{~\beta} + {\partial S^B({\vec q}, \tau) 
\over \partial q^\beta} \biggr) \biggr] \;.
\ee
A straightforward application of this formula is related to the 
Zel'dovich approximation. 
The spatial metric is that of Euclidean space in time--dependent 
curvilinear coordinates, consistently with the intuitive notion 
of Lagrangian picture in the Newtonian limit. 
Read this way, the complicated equations of Newtonian gravity in the 
Lagrangian picture become much easier: one just has to deal with the spatial 
metric tensor and its derivatives. 
The displacement vector is then completely fixed by solving the Raychaudhuri 
equation together with the momentum constraint in the $c \to \infty$ limit. 

Next, we can consider the post--Newtonian corrections to the metric and 
write equations for them. In particular, we can derive a 
simple and general equation for the gravitational--waves $\pi_{\alpha\beta}$
emitted by non--linear structures described through Newtonian gravity. The 
result can be expressed both in Lagrangian and Eulerian coordinates. In the 
latter case one has, 
\be
\nabla^2_x \pi_{AB} = \Psi^{(E)}_{v,AB} + \delta_{AB} \nabla_x^2 \Psi_v^{(E)}
+ 2 \biggl( \bar \vartheta \bar \vartheta_{AB} -
\bar \vartheta_{AC}
\bar \vartheta^C_{~~B} \biggr) \;,
\ee
with capital latin labels $A,B, \dots = 1,2,3$ indicating Eulerian 
coordinates and $\nabla_x^2 \Psi_v^{(E)} = - \frac{1}{2} 
( \bar \vartheta^2 - \bar \vartheta^A_{~B} \bar \vartheta^B_{~A} )$,
which generally allows a simple derivation of $\pi_{AB}$, given the
(gradients of the) velocity potential,
$\bar \vartheta_{AB} = \partial^2 \Phi_v/\partial x^A \partial x^B$, 
by a convolution in Fourier space.
These formulae would allow to calculate the 
amplitude of the gravitational--wave modes in terms of the velocity 
potential, which in turn can be deduced from observational data on 
radial peculiar velocities of galaxies. 

In the standard case, where the cosmological perturbations form 
a homogeneous and isotropic random field, we can obtain a heuristic 
perturbative estimate of their amplitude in terms of the 
{\em rms} density contrast and of the ratio of the typical perturbation scale 
$\lambda$ to the Hubble radius $r_H=c H^{-1}$. One simply has
$\pi_{rms} / c^2 \sim \delta_{rms}^2 (\lambda / r_H )^2$. 
This effect gives rise to a stochastic background of 
gravitational waves which gets a non--negligible amplitude in 
the so--called {\em extremely--low--frequency} band 
(e.g. Thorne 1995), around $10^{-14}$ --  $10^{-15}$ Hz. 
We can roughly estimate that the present--day closure density of this
gravitational--wave background is 
\be
\Omega_{gw}(\lambda) \sim \delta_{rms}^4 
\biggl( {\lambda \over r_H} \biggr)^2 \;.
\ee 
In standard scenarios for the formation of structure in the universe, 
the typical density contrast on scales 
$1$ -- $10$ Mpc implies that $\Omega_{gw}$ is about $10^{-5}$ -- 
$10^{-6}$. We might speculate that such a background would give rise to 
secondary CMB anisotropies on intermediate angular scales: a sort of 
{\em tensor Rees--Sciama effect}. This issue will be considered in 
more detail elsewhere. 

The previous PN formula also applies to isolated structures, where 
the density contrast can be much higher than the {\em rms} value, 
and shear anisotropies play a fundamental role. A calculation of 
$\pi_{\alpha\beta}$ in the case of a homogeneous ellipsoid 
showed that the PN tensor modes become 
dominant, compared to the Newtonian contributions to the metric tensor, 
during the late stages of collapse, and possibly even in a 
shell--crossing singularity. It is 
important to stress that this effect generally contradicts the standard 
paradigm that the smallest scale for the applicability of the 
Newtonian approximation is set by the Schwarzschild radius of the object. 
Such a critical scale is indeed only relevant for nearly spherical collapse, 
whereas this effect becomes important if the collapsing structure 
strongly deviates from sphericity.


\begin{references}
\reference Bertschinger E., Jain B., 1994, ApJ, 431, 486
\reference Bertschinger E., Hamilton A., 1994, ApJ, 435, 1
\reference Bruni M., Matarrese S., Pantano O., 1995a, ApJ, 445, 958
\reference Bruni M., Matarrese S., Pantano O., 1995b, Phys. Rev. Lett., 74, 11
\reference Buchert T., 1995, to appear in Proc. Enrico Fermi School, 
Course CXXXII, Dark Matter in the Universe, Varenna 1995, preprint 
astro-ph/9509005
\reference Catelan P., Lucchin F., Matarrese S., Moscardini L., 1995, MNRAS, 
276, 39 
\reference Futamase T., 1991, Prog. Theor. Phys., 86, 389
\reference Kasai M., 1995, Phys. Rev., D52, 5605
\reference Kofman L., Pogosyan D., 1995, ApJ, 442, 30
\reference Matarrese S., Pantano O., Saez D., 1993, Phys. Rev., D47, 1311
\reference Matarrese S., Pantano O., Saez D., 1994a, Phys. Rev. Lett., 72, 320
\reference Matarrese S., Pantano O., Saez D., 1994b, MNRAS, 271, 513
\reference Matarrese S., Terranova D., 1995, preprint astro-ph/9511093, 
submitted to MNRAS
\reference Peebles P.J.E., 1993, Principles of Physical Cosmology, 
Princeton University Press, Princeton
\reference Shandarin S.F., Melott A.L., McDavitt K., Pauls J.L., Tinker J., 
1995, Phys. Rev. Lett., 75, 1
\reference Shibata M., Asada H., 1995, Prog. Theor. Phys., 94, 11
\reference Thorne K.S., 1995, to appear in Kolb E.W. \& Peccei R., eds., Proc. 
Snowmass 95 Summer Study on Particle and Nuclear Astrophysics and Cosmology, 
World Scientific, Singapore, preprint gr-qc/9506086
\reference Tomita K., 1967, Prog. Theor. Phys., 37, 831
\reference Tomita K., 1991, Prog. Theor. Phys., 85, 1041

\end{references}
\end{document}